\documentclass[letterpaper, 10 pt,conference]{ieeeconf}
\IEEEoverridecommandlockouts
\usepackage{graphicx}
\usepackage{fixltx2e}
\usepackage{epsfig}
\usepackage{amsmath}
\usepackage{amsfonts}
\usepackage{amssymb}
\usepackage{latexsym}
\usepackage{multirow}
\usepackage{cite}

\usepackage{units}
\usepackage{url}
\usepackage{balance}
\usepackage{ucs}

\usepackage[german]{babel}

\usepackage{fancyhdr}
\title{ {\LARGE IAC-09.A6.4.11\\
MODEL OF AN INTERNATIONAL ENVIRONMENTAL AGREEMENT AMONG ASYMMETRIC NATIONS APPLIED TO DEBRIS MITIGATION}}
\author{ \parbox{3 in}{\centering Michael J. Singer, Graduate Student Researcher\\
John T. Musacchio, Assistant Professor
         \thanks{Supported by the University of California, Santa Cruz-NASA University Affiliated Research Center (UARC).}\\
         Technology and Information Management\\
         University of California, Santa Cruz\\
         {\tt\small \{mjsinger/johnm\}@soe.ucsc.edu } }
}

\begin{document}
\maketitle

\addtolength{\voffset}{-8mm}
\thispagestyle{fancy}
\lhead{60th International Astronautical Congress 12-16/10/2009 Daejeon, Korea\\Session A6.4 Space Debris: Mitigation and Standards}
\cfoot{}

\begin{abstract}
We investigate how ideas from the International Environmental Agreement (IEA) literature can be applied to the problem of space debris mitigation. The problem of space debris is similar to other international environmental problems in that there is a potential for a tragedy of the commons effect--individual nations bear all the cost of their mitigation measures but share only a fraction of the benefit. Consequently, nations have a tendency to underinvest in mitigation. Coalitions of nations, brought together by IEAs, have the potential to lessen the tragedy of the commons effect by pooling the costs and benefits of mitigation. This work brings together two recent modeling advances: i) a game theoretic model for studying the potential gains from IEA cooperation between nations with asymmetric costs and benefits, ii) an orbital debris model that gives the societal cost that specific actions, such as failing to deorbit an inactive satellite, have on the environment. We combine these two models with empirical launch share data for a ``proof of concept'' of an IEA for a single mitigation measure, deorbiting spacecraft at the end of operational lifetime. Simulations of all possible coalitions for a proxy set of 12 asymmetric nations suggest the possibility that stable coalitions can provide significant deorbiting gains relative to nations acting in the absence of an IEA coalition.
\end{abstract}

\section{\underline{Problem Statement}}

When actions of individuals affect a shared resource, there is potential for a tragedy of the commons scenario: individual decision-makers under-invest in protection if they see only a fraction of the benefits from the investment. Consequently, a stream of recent literature has sought to understand how nations can form coalitions to counter the potential for tragedy of the commons scenarios in protecting the environment.

Game-theoretic models in the greenhouse gas (GHG), ozone depletion, and acid rain arenas have shown that when nations 1) recognize asymmetries of marginal costs and benefits of mitigation and 2) establish coalitions to transfer payments to adjust abatement rates, there can be a substantial increase in global levels of pollution abatement. The size of the increase is a function of the nature and extent of the asymmetries\cite{matthew_mcginty_international_2007}.

Our initial focus applies the IEA framework to one debris mitigation measure, post-mission deorbiting of spacecraft, using marginal benefits derived from the damage metric provided by Bradley and Wein (2009)\cite{bradley_space_2009}, marginal costs derived from deorbit cost estimates given by Weidemann (2004)\cite{wiedemann_analyzing_2004}, and spacecraft ownership data from the Union of Concerned Scientist database\cite{union_of_concerned_scientists_ucs_2009}.

\section{\underline{IEA Model Framework}}

\subsection{\underline{Overview}}
Game-theoretic methods are ideally suited to formally modeling strategic considerations of actors causing transboundary environmental externalities\cite{finus_endogenous_2001}. 

The fundamental assumption of IEAs is that, given the lack of an international agency to enforce agreements, they must be designed to be self-enforcing. ``For IEAs to improve management of shared environmental resources, they must make it attractive for countries to sign and carry out the terms of the agreement''\cite{barrett_self-enforcing_1994}. The theory assumes parties act in their own self-interest to maximize their individual net profit (i.e., share of benefit from global abatement minus individual costs of environmental abatement). 

In game-theoretic framing, each actor independently attempts to find a ``best response'' to other players' strategies. If there is a mutual best response where no player can benefit by individually deviating from that solution, the game solution is a ``Nash-equilibrium'' \cite{fudenberg_game_1991}. 

In contrast with this ``Nash'' (here termed ``null coalition'') behavior, game-theoretic research in the IEA domain poses a ``social'' optimal or ``full-cooperative''\cite{matthew_mcginty_international_2007} behavior that maximizes global profit summed over all actors. All actors act in concert as a ``full coalition'' (also referred to in the literature as a grand or complete coalition), behaving as if dictated by a ``social hegemon'' decreeing individual abatement levels.

IEA research has focused on evaluating levels of abatement and profit for self-enforcing ``partial coalitions'' relative to null coalition and full coalition values under different assumptions of player characteristics and rules of coalition formation.

For a partial coalition to be stable (i.e., self-enforcing), three criteria must be satisfied:
\begin{enumerate}
\item Coalition members individually realize a greater benefit under the agreement than they would outside (a condition termed ``internal stability'' in economic oligopoly literature),
\item Non-members do not perceive an incentive to join (``external stability''), and
\item Total coalition profit exceeds total payoffs for each single defector (``coalition stability'')\cite{matthew_mcginty_international_2007}.
\end{enumerate}

\pagestyle{plain} 

IEA modeling has used both cooperative (i.e., with a mediator) and non-cooperative game-theoretic approaches \cite{finus_game_2007}, with seminal works including the non-cooperative  ``benchmark'' model introduced by Barrett \cite{barrett_self-enforcing_1994}, Hoel \cite{michael_hoel_international_1992}, and Carraro and Siniscalco \cite{carraro_strategies_1991}.
A key challenge for IEAs has been finding mechanisms to overcome incentives for nations to ``free-ride'' on the abatement efforts of others\cite{finus_endogenous_2001}.

Early research \cite{barrett_self-enforcing_1994} indicated that a partial coalition formed by identical nations (i.e., each nation having identical cost functions and benefit functions for abatement) could neither attain significant membership (no more than three members, depending on the shape of the marginal benefit and cost functions) nor improve significantly upon the null coalition level of global abatement. Later research by Barrett \cite{barrett_heterogeneous_1997}, generalized by McGinty\cite{matthew_mcginty_international_2007}, indicated potentially greater membership and benefits for self-enforcing coalitions relative to null coalition outcome when there are marginal cost and benefit asymmetries between actors (nations).

We adopt the non-cooperative game model of McGinty 2007\cite{matthew_mcginty_international_2007}, extending Barrett's 1997 and 2001 analyses\cite{barrett_heterogeneous_1997} and \cite{barrett_international_2001}, to investigate stability and effectiveness of self-enforcing partial coalitions in contrast with the null coalition and the full coalition.

The global profit function in all three cases is given in general form by:
\begin{equation}
\Pi={\sum}\pi_{i} \text{ where } \pi_i=B_i(Q)-C_i(q_i).
\end{equation}

While each nation bears the cost of its own abatement, $q_i$, all nations share the benefits of global abatement, $Q$.

\subsection{\underline{IEA Model (after McGinty (2007))}}
The profit function for each player is:

\begin{equation}
\pi_{i}=b\alpha_{i}\left(aQ-\frac{Q^{2}}{2}\right)-\frac{c_{i}q_{i}^{2}}{2}
\end{equation}

\noindent where $\alpha_{i}$ represents share of global benefits, parameters
$a$ and $b$ can be adjusted to reflect how quickly marginal benefits
decrease as abatement increases, and $c_{i}$ is the cost coefficient. Marginal benefits and costs in this model are linear and asymmetric.

McGinty's model computes, in addition to global and individual abatement and profit for null and full coalition, corresponding values for each partial coalition chosen from the power set of all possible coalitions. The model distinguishes abatement and profit for coalition members from that of non-members.

In a stable partial coalition, members have a collective profit that is at least as high as the sum of their individual profits operating alone. The most efficient allocation according to cost requires that all nations abate proportionately with the ratios of their share of global benefit to marginal cost coefficient. But the efficient level of abatement for a given member might not result in a higher payoff for an individual member compared to leaving the coalition.

To overcome this problem, the final step is designing a transfer payment scheme that optimizes net burden allocation among members (i.e., required share of total cost). 

Those members for whom the required share of abatement is greater than the abatement they perform pay into a pot; those for whom the required share is less than that performed get paid from the pot. Net transfers are zero-sum.

The one necessary requirement for this ``burden-sharing'' rule is that it has to be incentive-compatible\textemdash i.e.,
transfers ensure the profit each nation receives as a coalition member exceeds that they would outside the coalition. The ``burden sharing'' rule is otherwise flexible, allowing for optimization to maximize:
\begin{itemize}
\item Abatement effectiveness
\item Membership
\item Equity (on a per capita basis, to reflect legacy considerations, etc.)
\end{itemize}

McGinty proposes an allocation rule that is optimal in that ``transfers are just sufficient to quell any incentive to deviate from the agreement.'' The rule distributes the surplus that remains ``once each coalition member receives their payoff as a single defector from the IEA ... in proportion to their benefit [share]-cost ratio.'' (pp. 3, 6)

[We provide an overview of McGinty's treatment of the partial coalition case in the Appendix and refer the reader to \cite{matthew_mcginty_international_2007} for details.]

\section{\underline{Application to Debris Mitigation}}

As an initial exploration of modeling an IEA for space debris mitigation, we examine outcomes for parties choosing a single type of abatement action: whether to equip satellites with de-orbit capability. We ignore other actions parties might take to abate debris generation, remove objects from the environment, or mitigate the effects of debris (e.g., by increasing spacecraft shielding).

\subsection{\underline{Elements of the Model}}

An IEA model requires the following elements: 

\begin{itemize}
\item An environmental resource and a pollutant,
\item Actors (or, in game parlance, ``players''; here, actors are nations or groups of nations) and actions they do or can take to affect the environment, 
\item A model to quantify harm to the environment from pollutant generation,
\item A model to quantify benefits to actors for their own and others' 
abatement of harm, and
\item A model to quantify costs to each actor for abating harm.

\end{itemize}

We specify the elements as follows for a space debris mitigation IEA:

\subsubsection{\underline{Environmental Resource and Pollutant}}

The environmental resource analyzed here is near-earth space, in particular, the shell-of-interest (SOI) of high debris flux at 900-1000 km altitude in low-earth orbit (LEO) analyzed by Bradley and Wein\cite{bradley_space_2009}. The pollutant is a subset of orbital objects: future spacecraft launched but not deorbited after completion of their operational lifetimes. Potential for collision with other spacecraft or debris increases the risk of additional debris generation.

\subsubsection{\underline{Actors and Actions}}

\begin{figure}
     \begin{center}
        \includegraphics[height=2.5in]{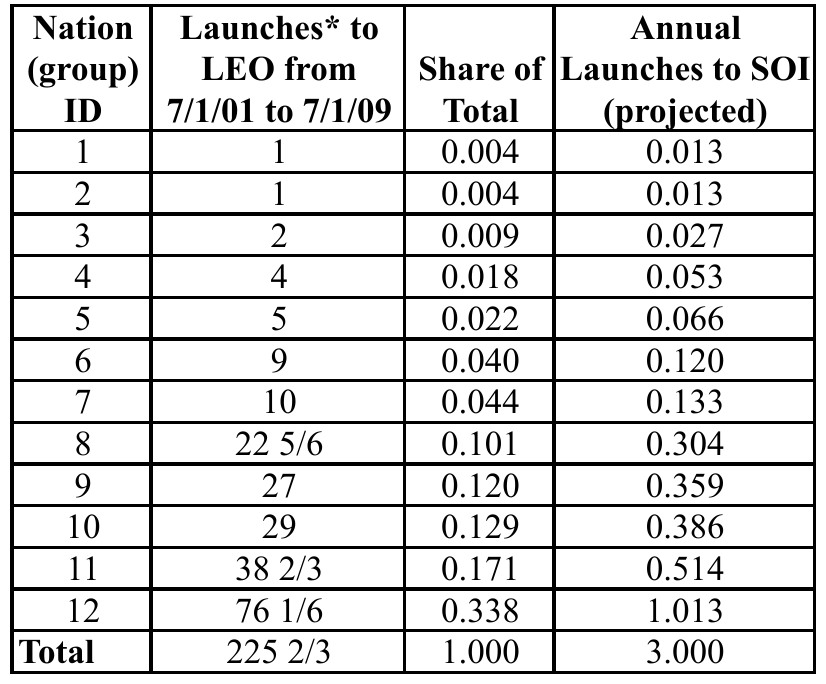}
        \end{center}
        \caption{Projected launch rate to the SOI for a proxy set of spacecraft-owning nations, sorted low to high (*reflecting shared ownership of some spacecraft) [derived, with liberties taken, from the UCS database]}
        \label{f:Launch_Rate}
\end{figure}

We recognize the difficulty in predicting the set of spacecraft owners and their launch rates to the SOI. Somewhat arbitrarily, we construct a proxy set of future spacecraft owners derived from nation-states and international organizations listed in the Union of Concerned Scientists database of operational spacecraft as launching spacecraft to low earth orbit (LEO). The nations are identified by their ranking from 1 (lowest launch rate) through 12 (highest launch rate). For this proxy set of 12 nations, Figure~\ref{f:Launch_Rate} shows 1) launch rates to LEO for a recent eight-year period, 2) share of total annual launches, and 3) annual launch rates to the SOI after scaling total launch rate to that taken by Bradley and Wein for this SOI (three operational spacecraft per year)\cite{bradley_space_2009}.\footnote{The proxy set reflects adjustments such as grouping ESA and all European nations into a single entity, taking only a sampling of nations with lower launch rates, and evenly dividing ownership for spacecraft listed as having multiple owners. The length of historical launch period follows practice for long-term debris models\cite{martin_sensitivity_2004,liou_leo_2005}.}

At this point of model development, we ignore issues arising from some nations' very low launch rates (e.g., one every 75 years for two listed nations) and assume these issues can be addressed by such mechanisms as ``issue linkage'' (e.g., linking IEAs for different SOIs), subcoalitions, ``pools'' of nations, or financial derivative instruments.

The only abatement action we consider in this preliminary analysis is deorbiting. We adopt Bradley and Wein's assumption that deorbiting from the SOI takes place instantaneously.

\subsubsection{\underline{Harm}}

Bradley and Wein (2009), in the exposition of their debris environment model, introduce several important performance metrics relevant to environmental risk assessment. The model, a mean-field approximation set of ordinary differential equations, computes rates of change of spacecraft, rocket bodies, and fragments in a SOI for $T \in [0,\infty)$, where 0 is the present. The model categorizes spacecraft as operational or no longer operational, with or without deorbit capability; rocket bodies with or without deorbit capability; and fragments as hazardous or benign in collision with other objects depending on collision velocity and fragment characteristics. Parameters for the differential equations are expectations over the same distributions that govern an object-by-object simulation. [We provide an overview of Bradley and Wein's 2009 model in Appendix B and refer the reader to \cite{bradley_space_2009} for details.]

\begin{figure}
     \begin{center}
        \includegraphics[height=2.5in]{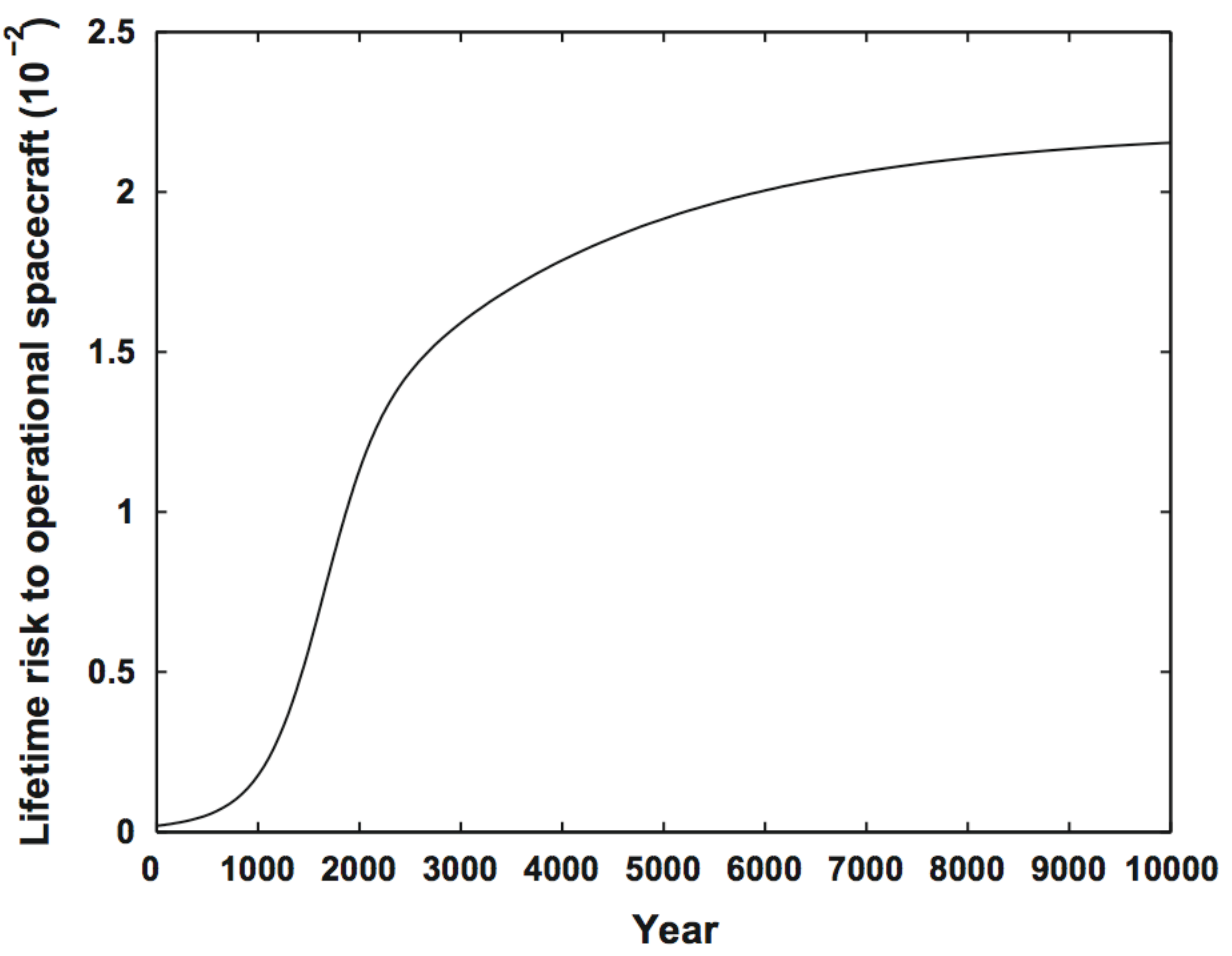}
        \end{center}
        \caption{Risk to a spacecraft of catastrophic destruction during its operational lifetime. [Bradley and Wein 2009, Fig. 4b, reprinted with permission]}
        \label{f:Bradley_Wein_2009_Fig_4b}
\end{figure}

The model's primary metric is ``lifetime risk'', ``the probability that a spacecraft launched at time $t$ will be destroyed (via an intact-intact or catastrophic intact-fragment collision) while it is still operational'' (p. 1376). Figure~\ref{f:Bradley_Wein_2009_Fig_4b} shows ``lifetime risk'' for an operational spacecraft launched at the present epoch in the SOI given a baseline set of parameters (launch rates, existing debris flux, spacecraft characteristics, fraction of spacecraft deorbited, etc.). Risk increases at a modest rate for the next several hundred years, increases rapidly starting at about year 1000, then levels off ca. 3000 years to approach an equilibrium value.\footnote{See p. 1381 of \cite{bradley_space_2009} for discussion of differences in long-term predictions given by this model and Kessler's 2000 model \cite{kessler_critical_2001} and differences regarding active debris removal conclusions relative to those drawn by Liou and Johnson \cite{liou_risks_2006} and \cite{liou_instability_2008}.}

\begin{figure}
     \begin{center}
        \includegraphics[height=2.5in]{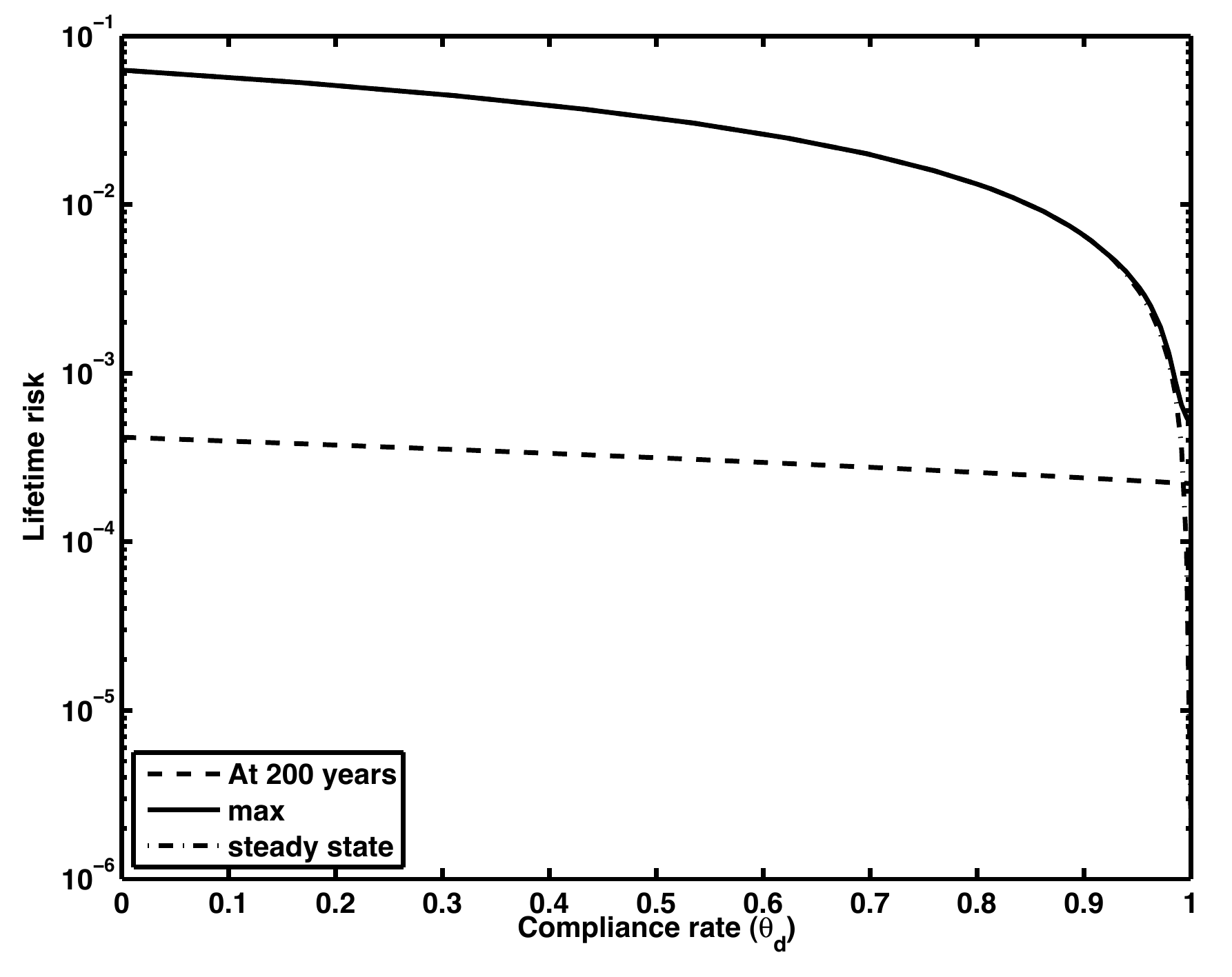}
        \end{center}
        \caption{Sustainable lifetime risk (solid curve) and risk at 200 years (dashed line) to an operational spacecraft as a function of fraction of new launches that deorbit the spacecraft at the end of its operational life. [Bradley and Wein 2009, Fig. 5, reprinted with permission]}
        \label{f:Bradley_Wein_2009_Fig_5}
\end{figure}

From the ``lifetime risk'' metric, Bradley and Wein  derive a second metric, ``sustainable lifetime risk'' defined as the maximum lifetime risk value over all future times. Figure~\ref{f:Bradley_Wein_2009_Fig_5} shows how the ``lifetime risk'' at 200 years (dashed line) and the ``sustainable lifetime risk'' (solid curve) vary as a function of fraction of spacecraft deorbited. The former value is relatively insensitive and the latter strongly sensitive to deorbit compliance.\footnote{Risk at steady state differs from the maximum (i.e., ``sustainable lifetime risk'') only at very high ($>97\%$) compliance rates; note that Figure~\ref{f:Bradley_Wein_2009_Fig_4b} shows risk still increasing 10,000 years from present for a  \nicefrac{2\,}{\,3} compliance rate. } 

\begin{figure}
     \begin{center}
        \includegraphics[height=2.5in]{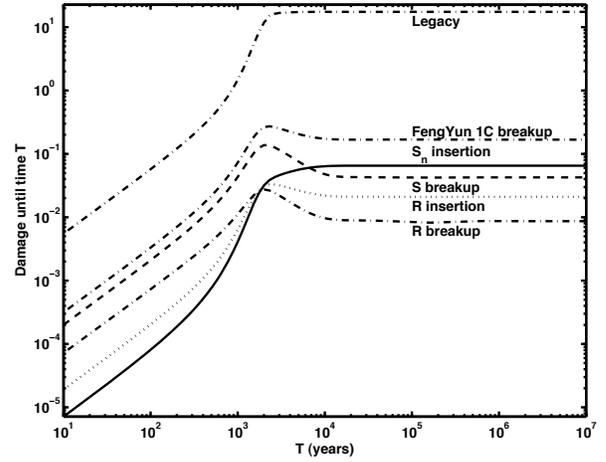}
        \end{center}
        \caption{Damage in terms of operational spacecraft destroyed up until time T caused by various activities that occur at time 0. Of key interest here is the ``$S_n$ insertion'' curve representing harm caused by failure to deorbit a spacecraft. (Bradley and Wein, 2009, p. 1379) [Bradley and Wein 2009, Fig. 6, reprinted with permission]}
        \label{f:Bradley_Wein_2009_Fig_6}
\end{figure}

From that second metric, Bradley and Wein  derive the measure of harm most directly relevant to our current purposes, the incremental number of operational spacecraft destroyed relative to baseline up until time $T$ caused by a failure to deorbit an ``extra'' spacecraft (i.e., a launch that represents a perturbation above baseline launch rate), shown as ``$S_n$ insertion'' in Figure~\ref{f:Bradley_Wein_2009_Fig_6}.\footnote{The subscript ``$n$'' signifies an ``extra'' non-operational spacecraft, i.e., a spacecraft not deorbited after operational lifetime; the loss of an ``extra'' satellite while still operational is not shown.} An interesting feature of the figure is that actions  representing break-ups at $T=0$ show the largest peak difference relative to baseline as the effects of that early breakup cause risk to rise sooner, more rapidly, and be of longer duration. Launch of a spacecraft or rocket body without deorbiting capability does not represent a current break-up but, instead, has an attendant risk of break-up at some future epoch. These curves, ``$R$ insertion'' and ``$S_n$ insertion'' evidence a smaller and no significant peak in damage, respectively, relative to baseline.

\subsubsection{\underline{Benefit from Abatement of Harm}}

\begin{figure}
     \begin{center}
\includegraphics[width=3in]{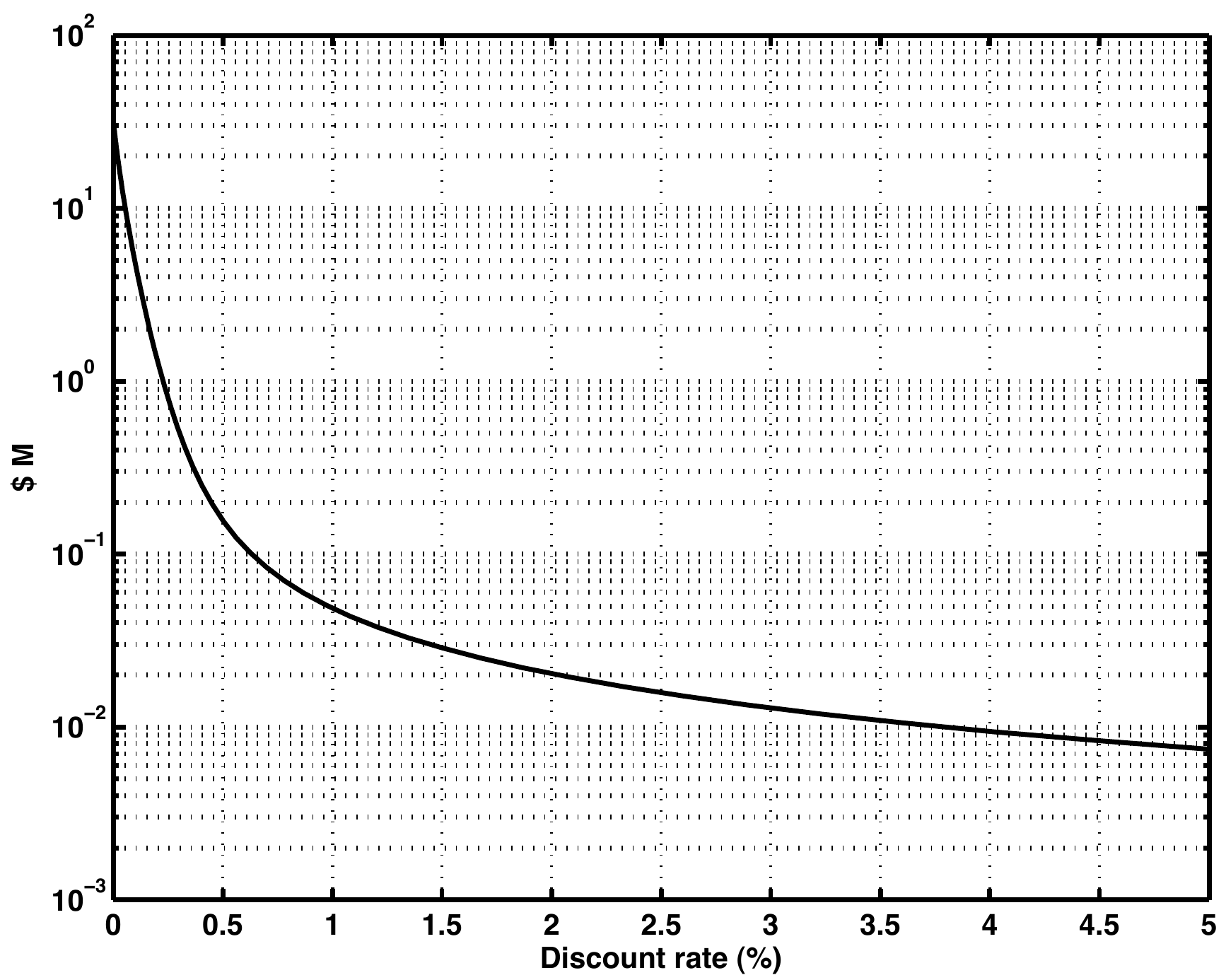}
        \end{center}
 \caption{Discounted damage from one satellite not being de-orbited at the end of its life, as a function of annual discount rate.  [personal communication from A. Bradley]}
  \label{f:discounted_damage}
\end{figure}

We adopt an average spacecraft value of \mbox{\$500M}, in accordance with Bradley and Wein. To determine benefit from abatement of harm by deorbiting a spacecraft, we use Bradley and Wein's third metric to calculate the net present value (NPV) of the incremental cost of operational spacecraft destroyed relative to baseline, as a function of discount rate (Figure~\ref{f:discounted_damage}): 

\[\text{NPV}_{\text{benefit}}=\text{\$500M}\int_0^{\infty} e^{-r t} \left[\frac{d({\text{damage}})}{dt}\right] dt\ .\] 

\medskip

Models in the GHG arena typically compute future harm for a period of 100 years and adopt discount rates on the order of 2\%\cite{dellink_stability_2008}. For convenience, we take as baseline a net present value of \mbox{\$1M} consistent with a discount rate of 0.25\%.\footnote{Such discount rates are not inconsistent with those recommended for valuing ``distant'' and ``far distant'' futures.\cite{weitzman_gamma_2001}} 

In our model, we set the product of benefit parameters $a\,b=$net present value of benefits from one deorbit=$1M$.\footnote{This value is chosen for illustrative purposes. Arguments could be made for either higher or lower valuations; future work will explore such schemes.} We assume benefit is a nearly linear function of compliance rate.\footnote{This assumption is based on Bradley and Wein Fig. 5.} To achieve this, we set the $b$ parameter to $10^{-8}$. 

We allocate abatement benefits to nations in accordance with their share of annual launches (Figure~\ref{f:Launch_Rate}). Nations with the greatest exposure (represented here only by their launch rates) to debris receive the greatest benefit from abatement.

\subsubsection{\underline{Abatement Costs}}

In keeping with the structure of the model of McGinty \cite{matthew_mcginty_international_2007} we assume the cost function for each nation $i$ deorbiting a quantity $q_i$ of satellites, is

\[ \frac{c_i q_i^2}{2} .\]

We suppose that a nation's marginal costs for deorbiting any one of its satellite is uniformly distributed between \$0 and \$1M. Deorbiting costs vary as a function of mass and other variables, in particular whether a satellite's mission already requires it to have maneuvering capabilities. Such a spacecraft might have a lower additional cost to add de-orbit capabilities than a satellite that would not otherwise need to carry fuel and thrusters for its mission. The \$0 to \$1M distribution is such that the average cost is \$0.5M.\footnote{This is the value Bradley and Wein extrapolate (p. 1378) from Weidemann, et al (2004) \cite{wiedemann_analyzing_2004} for the cost of deorbiting an ``average'' spacecraft (800 kg) from their shell-of-interest (900 to 1000 km).} 

We assume i) each nation chooses to add deorbit capability to those satellites for which addition is least costly and ii) the marginal cost of deorbiting increases linearly as the fraction of satellites a nation deorbits increases. Thus the marginal cost has the form \$1M $\times \frac{q_i}{n_i}$, and the total cost has the form \$1M $\times \frac{q_i^2}{2n_i}$. Note that, from the latter expression, the total cost to deorbit all $n_i$ of a nation~$i$'s satellites is 
\$0.5 $\times n_i$M, the average deorbit cost times the number of satellites.
Our assumption on the marginal cost is equivalent to assuming

\begin{equation}
c_i= \frac{\$1M}{n_i}\ .
\end{equation}

\subsection{\underline{Simulation}}

The simulation computes global and individual abatement and profit for the null and full coalition (social optimum) outcomes. It also computes, for each partial coalition chosen from the power set of all possible coalitions:

\begin{itemize}
\item Profit and abatement by members, and 
\item Profit by members if they were to leave the coalition.
\end{itemize}

The simulator then checks stability of each coalition. For each stable coalition, the simulation computes:

\begin{itemize}
\item Transfer payments among members (and member profits after transfer), and
\item Profit and abatement by non-members.
\end{itemize}

Finally, the simulator identifies those partial coalitions with the highest global abatement $Q$, profit $\Pi$, and number of members. 

\section{\underline{Results}}

\begin{figure}[h]
     \begin{center}
        \includegraphics[width=3.5in]{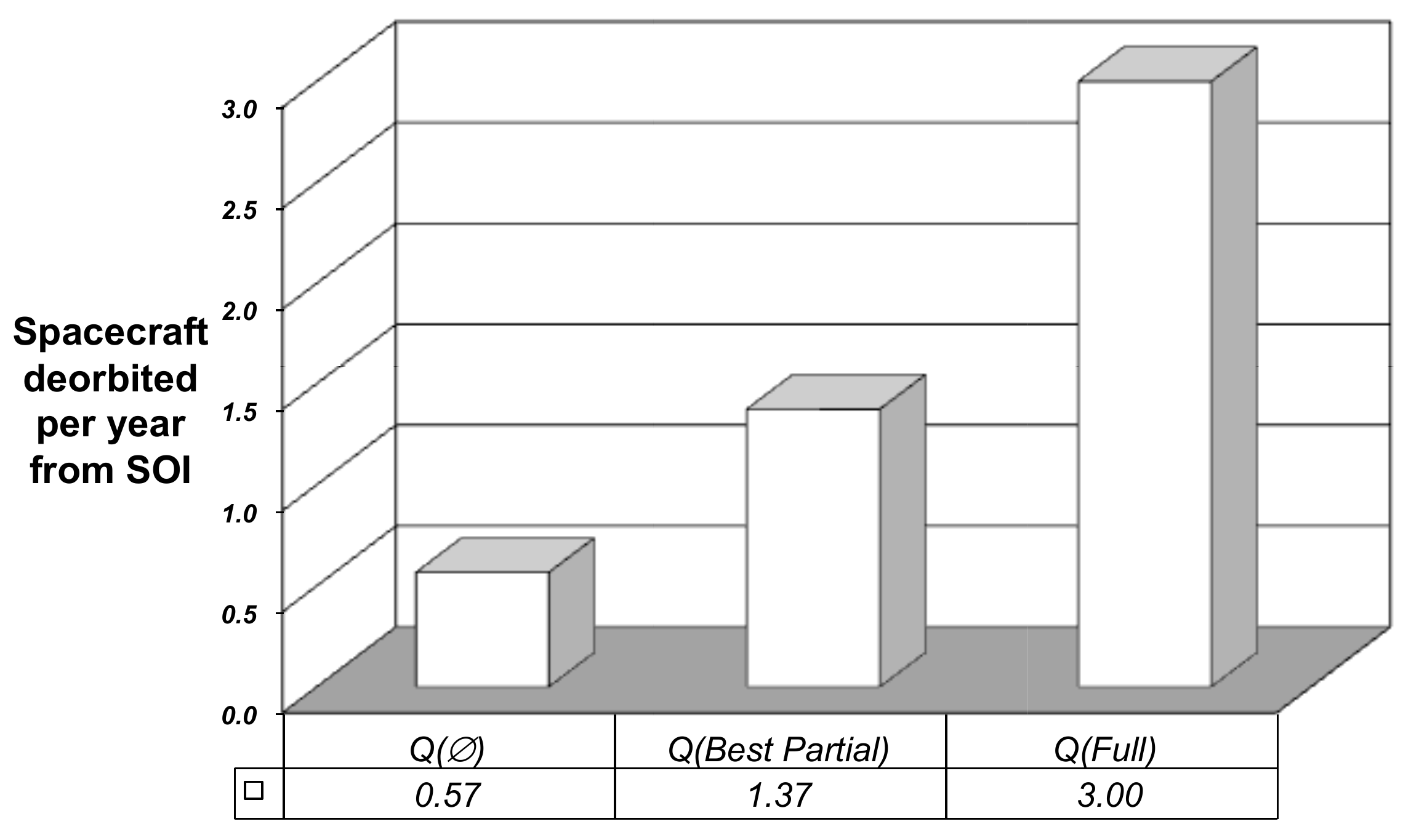}
        \end{center}
        \caption{Global abatement for null, best partial, and full coalitions}
        \label{f:globalQ}
\end{figure}

\begin{figure}
     \begin{center}
        \includegraphics[width=3.5in]{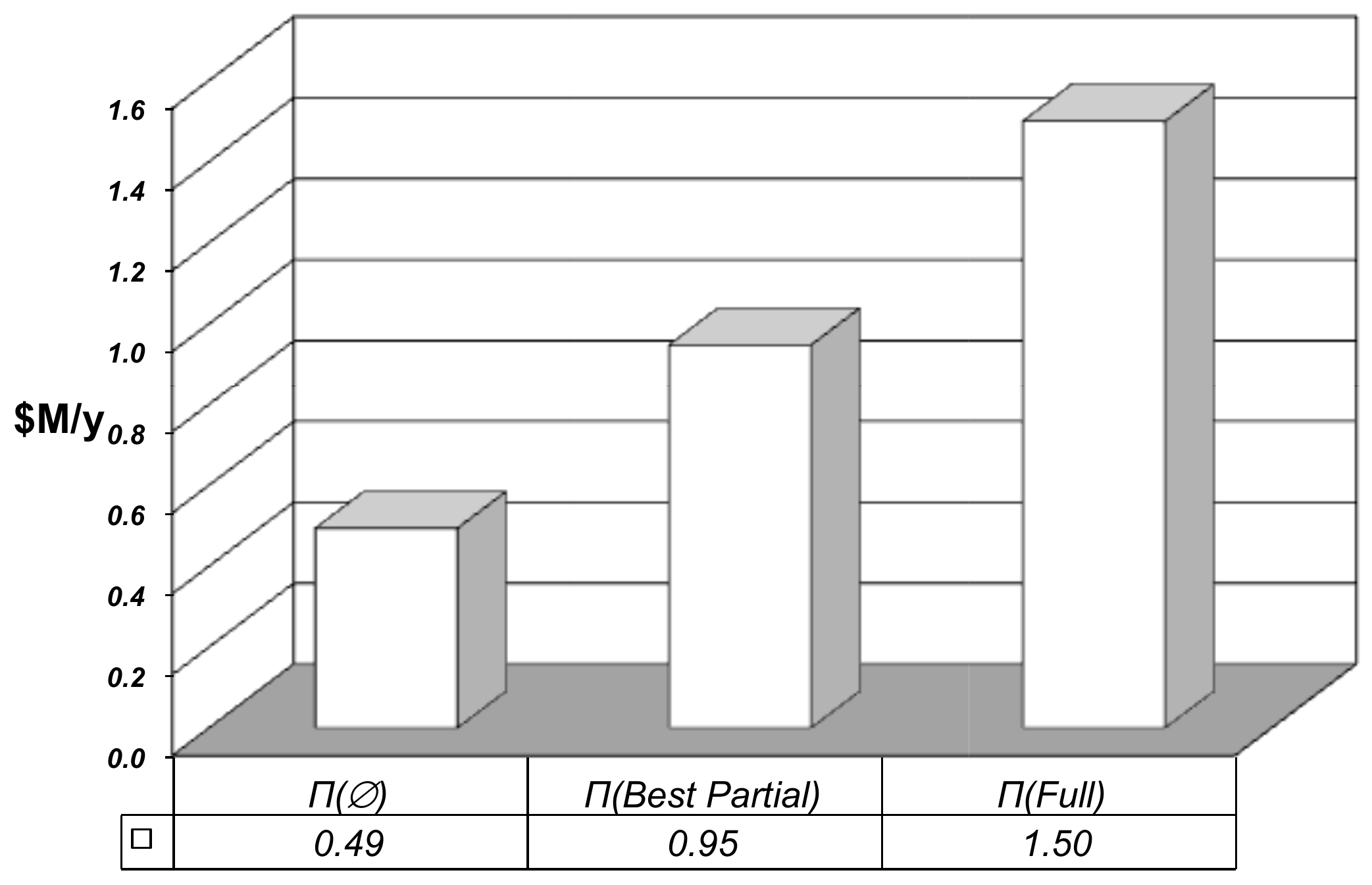}
        \end{center}
        \caption{Global profit for null, best partial, and full coalitions}
        \label{f:globalProfit}
\end{figure}

\begin{figure*}[h]
     \begin{center}
        \includegraphics[height=3in]{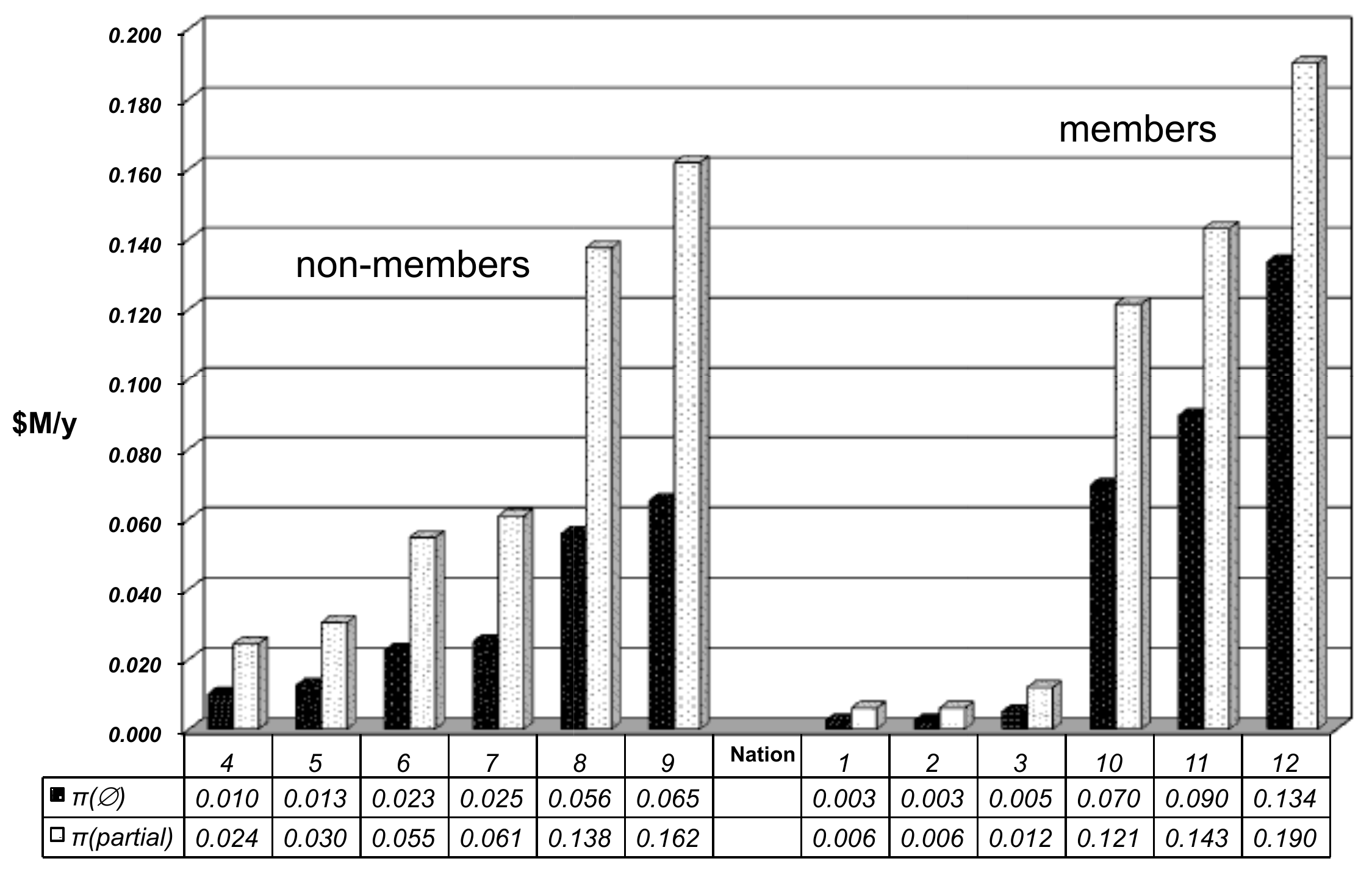}
        \end{center}
        \caption{Profit for null and best partial coalitions }
        \label{f:Profit_absolute}
\end{figure*}

\begin{figure*}[h]
     \begin{center}
        \includegraphics[height=3in]{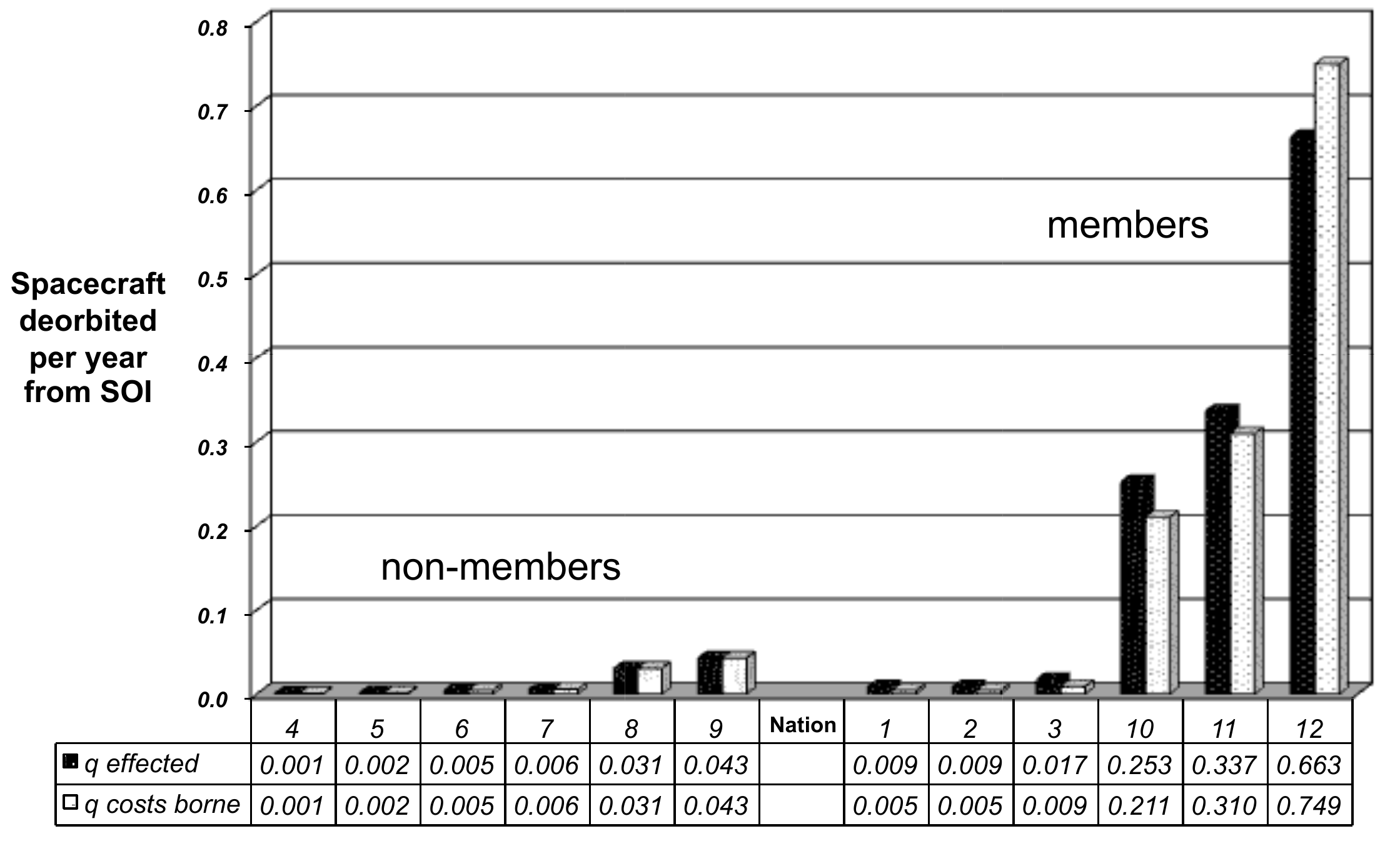}
        \end{center}
        \caption{Abatement effected vs abatement costs borne for best partial coalition }
        \label{f:qs_qr}
\end{figure*}

\begin{figure*}[h]
     \begin{center}
        \includegraphics[height=3in]{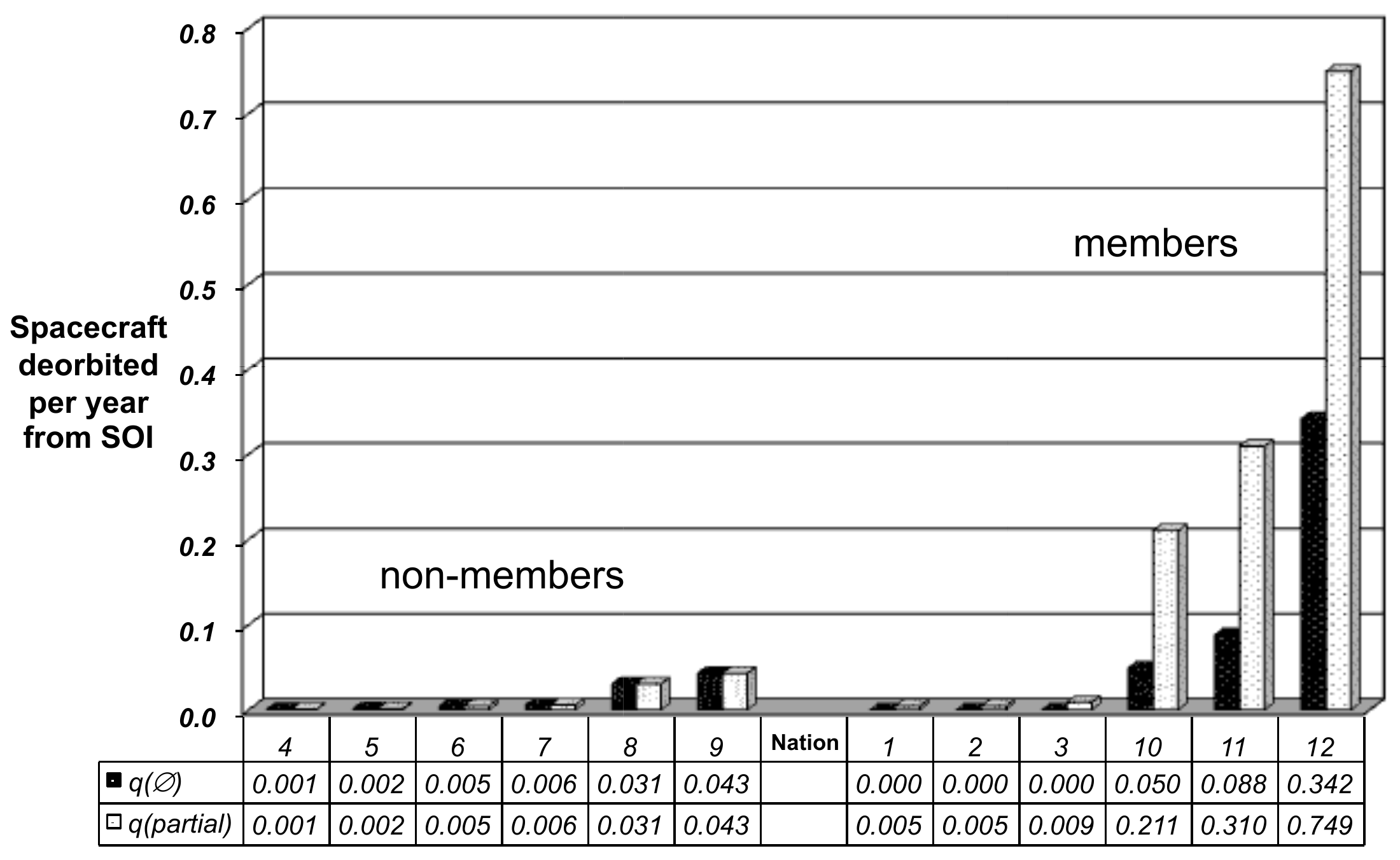}
        \end{center}
        \caption{Quantity of abatement: null vs best partial coalition}
        \label{f:q_absolute}
\end{figure*}

\begin{figure*}
     \begin{center}
        \includegraphics[height=3in]{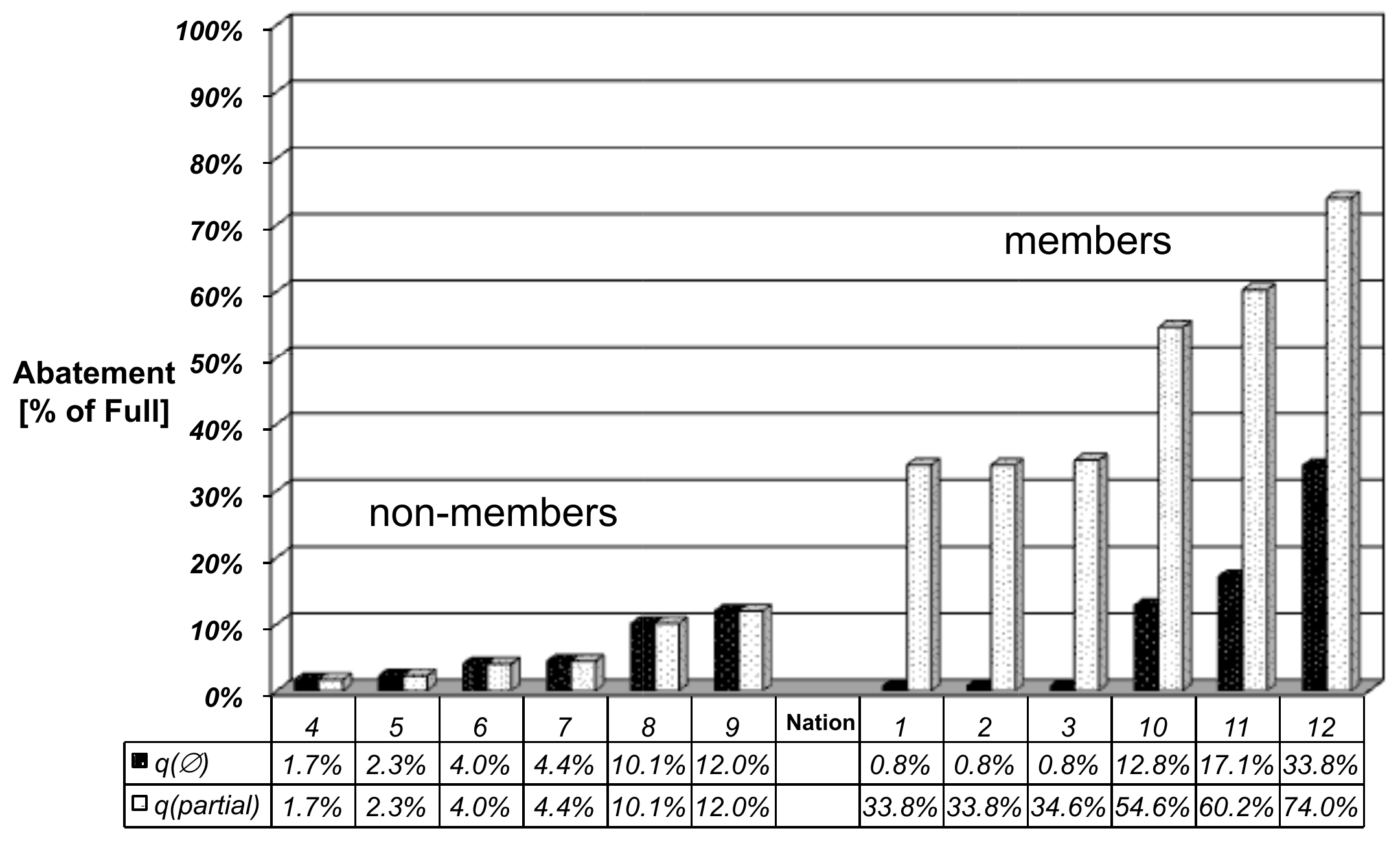}
        \end{center}
        \caption{Quantity of abatement: null vs best partial coalition as percent of full coalition}
        \label{f:q_percent}
\end{figure*}

For the simulation of twelve nations with parameterization as above\footnote{$a\,b = \$1\text{M}$, $b = 10^{-8}$, $\{\alpha_i\}$ from Figure~\ref{f:Launch_Rate}, $c_i=\frac{\text{\$1M}}{n_i}$}, there are 1280 stable coalitions out of $2^{12}=4096$ possible coalitions. The coalition that provides for the highest global abatement, \{1,2,3,10,11,12\}, is deemed the ``best partial coalition'' in the results below:

\begin{itemize}
\item Figure~\ref{f:globalQ} shows global abatement (deorbits/year) for the null, best partial, and full coalition cases. 
\item Figure~\ref{f:globalProfit} shows global profit (\$M/year) for the null, best partial, and full coalition cases. 
\item Figure~\ref{f:Profit_absolute} shows profit garnered by each nation for the null and best partial coalition cases.
\item Figure~\ref{f:qs_qr} shows abatement effected by each nation vs. abatement costs borne by each nation (after transfers) in the best partial coalition case.
\item Figure~\ref{f:q_absolute} shows abatement nation-by-nation for the null and best partial coalition cases, with non-members displayed on the left and members of the coalition displayed on the right.
\item Figure~\ref{f:q_percent} shows abatement nation-by-nation as a percentage of abatement each would be responsible for in the full coalition case.
\item Figure~\ref{f:Profit_percent} shows profit as a percentage of that each would be garner in the full coalition case.
\end{itemize}

\section{\underline{Discussion}}

The best partial coalition (the best stable coalition) achieves 46\% of the abatement of a full coalition (Figure 6). Recall that the full coalition case is equivalent to a  social optimum that could only be achieved if nations could be compelled to do what was socially optimal).  Moreover, the best partial coalition achieves 63\% of the payoff that a full coalition would achieve (Figure 7). 
In this case, the partial coalition that achieves the highest global abatement is the same coalition that achieves the highest global profit. (There are parameter settings for which this is not the case.) However, the best partial coalition is not the coalition with the largest membership--there are 7 and 8-member coalitions that do not perform as well on the other metrics. (The largest coalition, \{1, 2, 3, 4, 5, 6, 7, 12\}, executes only 30\% of the full coalition abatement.)

\begin{figure*}
     \begin{center}
        \includegraphics[height=3in]{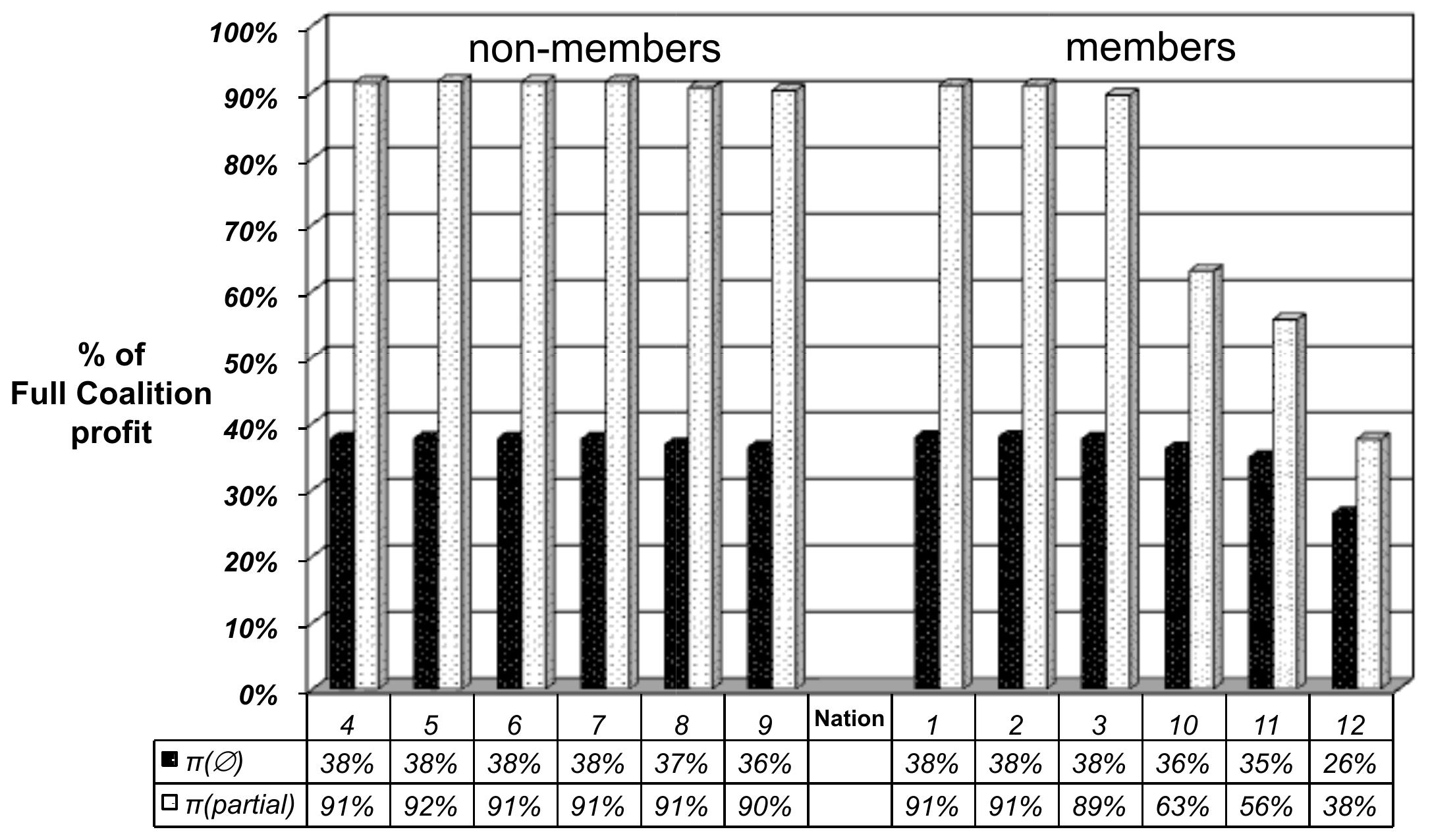}
        \end{center}
        \caption{Profit: null vs partial coalition as percent of full coalition}
        \label{f:Profit_percent}
\end{figure*}

Each member nation achieves a higher profit in the best partial coalition than they would in the null coalition case (Figures~\ref{f:Profit_absolute},~\ref{f:Profit_percent}). This confirms that membership in the coalition is consistent with parties' self-interest (i.e., membership is ``incentive-compatible'').

These profits incorporate transfers between nations. Figure~\ref{f:qs_qr} shows abatement effected by each coalition member in contrast with abatement costs borne by each member. The difference between these two quantities is the value of ``permits'' the member sells (or buys) within the coalition. In this case, nation 12 (which has the highest exposure to debris risk and, therefore, gains the most from debris abatement) pays all the other coalition members to increase their abatement from the levels they would execute according to their individual cost-benefit analyses.

As stated earlier, $a$ is a scaling parameter. Considering the profit function (Eq. 2) in view of the partial coalition solution for $Q$ given in Appendix A, one sees that $Q \propto a$ and $\Pi \propto a^2$. (The same proportionality holds for the null and full coalition cases.) Therefore, for any fixed value of $b$, the scaling derives, in turn, from the product $a\,b=\text{net present value of benefits from one deorbit}$: 
\begin{eqnarray*}
Q & \propto & \text{{(NPV\ of\ benefits\ from\ one\ deorbit)}}\\
\Pi & \propto & \text{{(NPV\ of\ benefits\ from\ one\ deorbit)}}^{2}\end{eqnarray*}

In the current model, we set the parameter $b$ to the arbitrarily small value of $10^{-8}$ in order to achieve a nearly linear benefit function. 

As can be seen in Figures~\ref{f:q_absolute} and ~\ref{f:q_percent}, all members of the best partial coalition significantly increase abatement over the levels they would execute in the absence of any coalition. Non-members do not reduce their abatement in the presence of the increased abatement by the coalition.

The independence of abatement behavior by non-members in our nearly linear case is consistent with prior explorations of linear benefit function. Consistent with prior research, as noted by McGinty\cite{matthew_mcginty_international_2007}, linear benefit functions result in orthogonal reaction functions for non-members, so IEA abatement does not influence non-member abatement (citing\cite{michael_hoel_international_1992}, \cite{barrett_heterogeneous_1997}, \cite{maler_acid_1989}).

The results suggest that, in a debris IEA where benefits are effectively additive in the regime of interest, active ``free-riding'' in the sense of a counter-productive decrease of non-members abatement to offset an increase in members' abatement may not be a problem. However, ``free-sharing'' of ``undeserved'' benefits to non-coalition members may be some psychological impediment to coalition formation. The extent to which non-members share in coalition benefits can be seen in Figure~\ref{f:Profit_percent}.

This work presents only a preliminary assessment of an IEA framework for debris mitigation. A number of issues remain to be explored to determine appropriateness of this approach given the special circumstances of the debris pollution problem:
\begin{itemize}
\item Explore discretization of abatement actions; variability of launch rates; periods of inactivity for selected players; variability of player sets; heterogeneity of pollution effect; small sample sizes; etc.
\item Consider linking IEAs for a set of orbital regimes or other issues;
\item Explore alternative benefit evaluation schemes;
\item Elaborate cost dependence of deorbiting with respect to spacecraft mass, orbit, and maneuvering system design parameters (e.g., prior inclusion of station-keeping, attitude control, collision avoidance capability); 
\item Incorporate other debris mitigation measures in strategy space: collision avoidance, solid rocket motor (SRM) slag prevention, active debris removal, etc.;
\item Account for non-catastrophic collisions that disable collision avoidance or deorbiting capability and for addition of shielding to protect these capabilities;
\item Reflect plausible distributions for launch rates, orbits, masses, and cross-sections for future launches;
\item Address broader set of stakeholder interests: military, government (non-military), civil, and commercial; non-spacecraft-owning nations; rogue activity, etc.;
\item Incorporate, as benefits, reduced damage to dependent systems (e.g., national infrastructure, improved national security);
\item Explore uncertainty and learning impacts on coalition performance;
\item Evaluate access to shared Space Situational Awareness (SSA) collision avoidance resources as an inducement providing direct benefit for coalition members and indirect benefit of protecting the space resource.
\end{itemize}

\section{\underline{Conclusions and Future Work}}

This simulation was a successful proof of concept of an International Environmental Agreement for debris mitigation. The results suggest that a coordination mechanism allowing for transfer payments between self-interested parties can provide a promising means for increasing abatement of debris generation. Future work will focus on verifying and extending these results and addressing the issues raised above.

\section*{\underline{Acknowledgements}}

The authors gratefully acknowledge Prof. Daniel Friedman, Dr. Matthew McGinty, and Mr. Andrew Bradley for their guidance, generosity, and valuable inputs to this work.

\medskip

\medskip

Figures \ref{f:Bradley_Wein_2009_Fig_4b}, ~\ref{f:Bradley_Wein_2009_Fig_5}, and ~\ref{f:Bradley_Wein_2009_Fig_6} reprinted from A. M. Bradley and L. M. Wein, ÒSpace debris: Assessing risk and responsibility,Ó \underline{Advances in Space Research}, vol. 43, pp. 1372-1390, May 2009, with permission from Elsevier.

\newpage

\newpage

\section*{\underline{Appendix A: After McGinty 2007\cite{matthew_mcginty_international_2007}}}

For each partial coalition, ${\bf {k}}$, chosen from the power set
of all actors, the global abatement is $Q\left({\bf {k}}\right)=Q_{s}\left({\bf {k}}\right)+Q_{t}\left({\bf {k}}\right)$. We determine coalition abatement 

\[
Q_{s}\left({\bf {k}}\right)=\frac{ab}{d}\ \sum_{i\in{\bf {k}}}\frac{1}{c_{i}}\underset{i\in{\bf {k}}}{\sum}\alpha_{i}\]

\noindent by maximizing members' joint profit, where
\[d=\left[1+b\sum{}_{j\notin{\bf {k}}}\theta_{j}\right]^{2}+b\sum_{i\in{\bf {k}}}\frac{1}{c_{i}}\sum_{i\in{\bf {k}}}\alpha_{i}\ ,\]

\medskip

Individual member (aka signatory) abatement

\[
q_{s}\left({\bf {k}}\right)=\frac{ab}{c_{s}d}\sum_{i\in{\bf {k}}}\alpha_{i}\]

\noindent is the most efficient allocation according to cost.

\medskip

Aggregate non-member abatement is \[
Q_{t}\left({\bf {k}}\right)=\frac{ab}{d}\sum_{i\in{\bf {k}}}\theta_{j}\left(1+b\sum{}_{j\notin{\bf {k}}}\theta_{j}\right)\]

\noindent and individual non-member abatement is \[
q_{t}\left({\bf {k}}\right)=\frac{ab}{d}\theta_{t}\left(1+b\sum{}_{j\notin{\bf {k}}}\theta_{j}\right).\]

\newpage
\section*{\underline{Appendix B: From Bradley and Wein 2009\cite{bradley_space_2009}}}

The rates of change for rocket bodies (i.e., upper stages, since first stages are assumed to deorbit immediately), spacecraft with deorbit capability, spacecraft that are still but which do not have deorbit capability, spacecraft that are no longer operational, and fragments of type $\kappa$ (hazardous or benign) from source $\tau$ (rocket body or spacecraft) are given by:

\begin{eqnarray*}
\dot{R}(t)=\lambda_{R}-\underset{\alpha\in U^{h}}{\sum}\beta_{R\alpha}R(t)\alpha(t)-\mu_{R}R(t)\\
\dot{S_{n}^{o}}(t)=(1-\theta_{d})\lambda_{o}-\underset{\alpha\in U^{h}}{\sum}\beta_{S\alpha}S_{n}^{o}(t)\alpha(t)-\mu_{o}S_{n}^{o}(t)\\
\dot{S_{n}}(t)=\mu{}_{o}S_{n}^{o}(t)-\underset{\alpha\in U^{h}}{\sum}\beta_{S\alpha}S_{n}(t)\alpha(t)-\mu_{n}S_{n}(t)\\
\dot{S_{d}}(t)=\theta_{d}\lambda_{o}-\underset{\alpha\in U^{h}}{\sum}\beta_{S\alpha}S_{d}(t)\alpha(t)-\mu_{o}S_{d}(t)\\
\dot{F_{\tau}^{\kappa}}(t)=\frac{1}{2}\underset{\alpha\in U}{\sum}\,\underset{\gamma\in U}{\sum}\delta_{\alpha\gamma}^{\tau\kappa}\alpha\left(t\right)\gamma\left(t\right)-\mu_{F_{\tau}^{\kappa}}F_{\tau}^{\kappa}(t)\end{eqnarray*}

\noindent where

\begin{eqnarray*}
\lambda & \equiv & \textrm{launch rate}\\
\theta & \equiv & \textrm{deorbit rate}\\
\mu^{-1} & \equiv & \textrm{average operational lifetime}\\
\beta, \delta & \equiv & \text{satellite and fragment collision rate parameters}\\
\alpha, \gamma & \in & \text{satellite types } \{S,R,F_{\tau}^{\kappa} \}
\end{eqnarray*}

Parameters for the differential equations are expectations over the same distributions that govern an object-by-object simulation. 

\newpage

\balance
\bibliographystyle{IEEEtranMod}
\bibliography{IAC_references_06a}

\end{document}